\documentclass{article}

\usepackage[nonatbib,final]{neurips_2021_ml4ps}

\usepackage[
    style=phys,
    sorting=none,
    maxcitenames=2,
    maxbibnames=10,
    doi=false,
    isbn=false,
    url=false,
    natbib=true,
    firstinits=true,
    hyperref,
    backend=bibtex]{biblatex} 
    
\AtEveryBibitem{%
  \clearfield{note}%
  \clearlist{language}
}

\usepackage[utf8]{inputenc} 
\usepackage[T1]{fontenc}    
\usepackage{hyperref}       
\usepackage{url}            
\usepackage{booktabs}       
\usepackage{amsfonts}       
\usepackage{nicefrac}       
\usepackage{microtype}      
\usepackage{graphicx}
\usepackage{amsmath,amssymb,amsthm}

\newtheorem{thm}{Theorem}
\newtheorem{prop}[thm]{Proposition}

\title{Cooperative multi-agent reinforcement learning for high-dimensional nonequilibrium control}

\author{Shriram Chennakesavalu \\
  Department of Chemistry \\
  Stanford University \\
  \texttt{shriramc@stanford.edu} \\
   \And
   Grant M. Rotskoff \\
   Department of Chemistry \\
   Stanford University \\
   \texttt{rotskoff@stanford.edu}
}

\usepackage[x11names, rgb, html]{xcolor}

\definecolor{green}{HTML}{0f6852}

\usepackage{hyperref}
\hypersetup{
  colorlinks = true,
  linkcolor = green,
  citecolor = green,
  urlcolor = teal
}

\begin{document}

\maketitle

\begin{abstract}
Experimental advances enabling high-resolution external control create new opportunities to produce materials with exotic properties. In this work, we investigate how a multi-agent reinforcement learning approach can be used to design external control protocols for self-assembly. We find that a fully decentralized approach performs remarkably well even with a "coarse" level of external control. More importantly, we see that a partially decentralized approach, where we include information about the local environment allows us to better control our system towards some target distribution. We explain this by analyzing our approach as a partially-observed Markov decision process.  
With a partially decentralized approach, the agent is able to act more presciently, both by preventing the formation of undesirable structures and by better stabilizing target structures as compared to a fully decentralized approach.
\end{abstract}

\section{Context and related work}
\label{intro}

Designing nanoscale structures that are tuned to have specific material properties or dynamical behavior is a longstanding goal in the molecular sciences~\cite{yin_colloidal_2005,ma_inverse_2019,gadelrab_inverting_2017, ronellenfitsch_inverse_2019}.
While advances in nanofabrication allow for increasingly intricate design, these approaches are often limited to specific materials and require homogeneity of the molecular components. 
Many nanoscale systems, however, ``self-assemble'' in suitable conditions or in the presence of appropriate external driving forces~\cite{cameron_biogenesis_2013,sigl_programmable_2021,rotskoff_robust_2018}. 
Exploiting self-assembly to manufacture nanoscale components presents its own challenges, as the system must be designed to spontaneously and rapidly form a given target structure.
The starting materials can become kinetically trapped, forming unfavorable structures if left to self-assemble without external control~\cite{rotskoff_robust_2018,whitelam_statistical_2015}.
Here we examine the problem of guiding self-assembly to target high yield of the target structure via external nonequilibrium driving forces, which we refer to as a ``protocol''. 
Reinforcement learning (RL) offers a promising toolkit for computing optimal protocols, but there has been little systematic investigation of approaches based on RL in this context.

Here, we consider two minimal models of molecular self-assembly in which particles evolve according to a stochastic, nonequilibrium dynamics.
In both models, clusters of particles form spontaneously under appropriate external conditions, and we seek to design protocols that optimize the size of particle clusters to a given target.
The external driving forces (e.g., temperature, light intensity) that we consider can be modulated as a function of both space and time (Fig. \ref{fig:active}).
In the systems considered here, the protocol is consequently high-dimensional because the external field is modulated on a grid with high spatial resolution. 
Because of the exponentially large action space, it is natural to employ multi-agent reinforcement learning (MARL) in this context.
Briefly, MARL extends the RL paradigm of an agent interacting with an environment to one in which multiple agents interact with the same environment concurrently (and possibly with each other).

Several works have investigated reinforcement learning to control condensed phase dynamics, mostly in equilibrium settings~\cite{rechtsman_optimized_2005, chen_patchy_18, romano_designing_2020, Ma_magnetic_self_assembly_2020}.
These approaches modulate the inter-particle interaction potentials between constituents---fundamentally altering the constituent material---rather than externally modulating them as we do here.
Recently, Ref.~\cite{falk_learning_2021} investigated controlling dynamics of a nonequilibrium active matter system with a low-dimensional protocol using standard RL techniques.
Multi-agent reinforcement learning has a substantial literature~\cite{tan_multi-agent_1993, busoniu_comprehensive_2008}; strategies that treat each cooperative agent independently are among the most relevant to the current study, including~\cite{rashid_qmix_2018, son_qtran_2019, sunehag_value-decomposition_2018}.

\paragraph{Main contributions:}
\begin{itemize}
    \item We argue that within the framework of a partially observed Markov Decision Process, multi-agent reinforcement learning with improved state estimates lead to improved value. 
    \item We demonstrate the effectiveness of incorporating local state and/or reward information in a MARL approach for controlling self-assembly in a system of active, nonequilibrium colloidal particles and equilibrium Lennard-Jones particles ungoing thermal annealing.
    
\end{itemize}

\section{Multi-Agent Reinforcement Learning}
\label{gen_inst}
We consider a partially observed Markov decision process (POMDP), cf. Ref.~\cite{roy_finding_2005}, specified as a tuple consisting of the state space, the action space, the space of observations, the observable function, the reward, the state-action transition operator, the discount factor, and the belief update:
$(\mathcal{S}, \mathcal{A}, \mathcal{Z}, \mathcal{O}, \mathcal{R}, T, \gamma, b_0).$
As depicted in Fig. 1, we consider an external control function that can independently tune the strength of an external field on a specific region.
We want to dynamically assemble clusters with a user-specified cluster size distribution $\rho_*$ and we take 
\begin{equation}
    \mathcal{C}(x) = D_{\rm KL}(\hat \rho(h(x)) \| \rho_*).
    \label{eq:reward}
 \end{equation}
 as our cost (or inverse reward) function. We discuss in detail how we compute this reward in ~\nameref{p:comp}.
 
 The function $h:\mathbb{R}^d \to \mathbb{N}^n$ takes a local particle configuration and simply counts the number of clusters with $k$ members for each $k<n$; this constitutes the observable $O.$ The empirical distribution of cluster sizes is denoted $\hat\rho.$ 
 Due to this mapping, the state information is not perfectly resolved by the agent, but rather the input information is \emph{partially observed}.
 We decompose the system into a fixed number of independent regions, each of which is controlled by an individual agent. 
Because each agent shares the same goal, the optimization is ``fully cooperative''~\cite{tan_multi-agent_1993}.
As a result, all agents share a common cost function~\eqref{eq:reward} and $Q$ function, allowing for high data efficiency as the experience of each agent is shared with all other agents.

This algorithm follows the MARL paradigm of centralized training but \emph{fully} decentralized execution. 
However, fully decentralized execution poses a limitation as spatially proximal information is crucial for the self-assembly of clusters. 
Because of the high spatial resolution of control, the presence or absence of particles and clusters in neighboring grids will no doubt influence the cluster formation for a grid.
Incorporating more local information into the observation of the current state leads to improved belief about the current state, which we summarize in the following proposition:
\begin{prop}[Low entropy belief improves expected reward]
\label{prop:belief}
Consider a POMDP given by $(\mathcal{S}, \mathcal{A}, \mathcal{Z}, \mathcal{O}, \mathcal{R}, T, \gamma, b_0)$ and the POMDP augmented with local state information in which the observation function is extended to $\tilde{\mathcal{O}}: \otimes_{i=1}^K \mathcal{S}_i \to \mathcal{Z}$ leading to a distinct evolution of the belief probabilities, denoted $\tilde b$.
Let $\bar s$ denote a state of the system with a unique optimal action $\bar a$. If $\tilde{\mathcal{O}}$ is such that $b(\bar s)=\delta_{s,\bar s}$ for all states $\bar s$, i.e. it perfectly resolves the state, then $V^*(\tilde b) \geq V^*(b)$ for all $\mathcal{O}\neq \tilde{\mathcal{O}}.$
\end{prop}
To prove this proposition, we simply write the Bellman optimality equation assuming the current state is $\bar s$, for which we know $\bar a$ attains the maximal value, and letting $\tau$ denote the transition operator on the belief vector induced by $T$,
\begin{equation}
\begin{aligned}
    V^*(\tilde{b}) &= \max_a \left[ \sum_{s} R(s,a) \tilde{b}(s) + \gamma \sum_{s'} p(\tilde{z} |\tilde{b},a) V^*(\tau) \right] \\
    &= \max_a \left[ R(\bar s, a) + \gamma \sum_{s'} T(s'|\bar s, a) V^*(s') \right],
\end{aligned}
\end{equation}
which is equivalent to the fully-observed MDP value function and $\tilde{z} = \tilde{\mathcal{O}}(s')$.
If the observation increases the uncertainty of the state relative to the perfectly resolved belief vector $\tilde b$, then there exists at least one state $s\neq \bar s$ that has belief $p_s>0.$
In other words, 
\begin{equation}
\begin{aligned}
    V^*(b) &= \max_a \left[ p_s R(s,a) + (1-p_s) R(\bar s, a) + \gamma \sum_{s'} p(\tilde{z}|\tilde{b},a) V^*(\tau) \right] \leq V^*(\tilde b),
\end{aligned}
\end{equation}
because the expected reward for any state $s\neq \bar s$ is suboptimal by assumption.

\paragraph{Computational approach\label{p:comp}}: We utilize a fully cooperative multi-agent Clipped Deep $Q$ Learning approach \cite{mnih-atari-2013, fujimoto_td3_18} to learn optimized protocols that minimize $\mathcal{C}$. For a given region consisting of particle configurations $x$, we define the state $s = h(x)$ and the cost as $\mathcal{C}(x)$. Additionally, we consider a discrete action space $\mathcal{A}$, representing the strength of the externally modulating field on that region. Proposition~\ref{prop:belief} suggests that incorporating additional local information should benefit the optimization, so long as that information improves the accuracy of the belief about the current state. To do so, we consider a hierarchy of local region sizes: a plaquette consisting of the current region and its nearest neighbors, a $3 \times 3$ grid, a $5 \times 5$ grid and the global system. Given particle configurations in the local regions, $x^\prime$, we define the state of the local region $s^\prime = h(x^\prime)$ and the cost of the local region as $\mathcal{C}(x^\prime)$. 

With this in mind, we extend our approach to include centralized training but \emph{partially} decentralized execution. This corresponds to sharing information among the agents, as we include information about the state and/or cost of local regions. When including state information about the local region, we consider the joint state between a region $s$ and its surrounding local region $s^\prime$. Additionally, when including cost information about the surrounding region, we consider the average cost between 
$\mathcal{C}(x)$ and $\mathcal{C}(x^\prime)$. We indicate whether or not we included surrounding state and/or cost information with a subscript S for cooperative state and a subscript C for cooperative cost.
\section{Active Colloids}

\begin{figure}[ht]
  \hfill\includegraphics[width=0.9\linewidth]{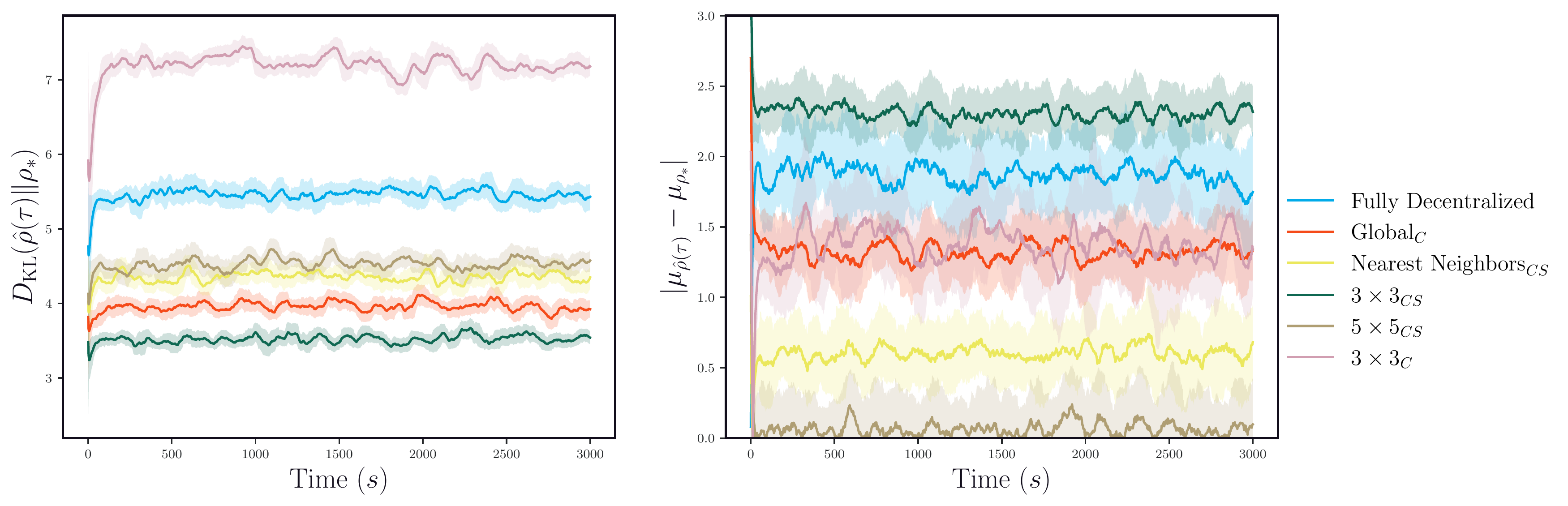}
  \begin{center}
  \includegraphics[width=0.2\linewidth]{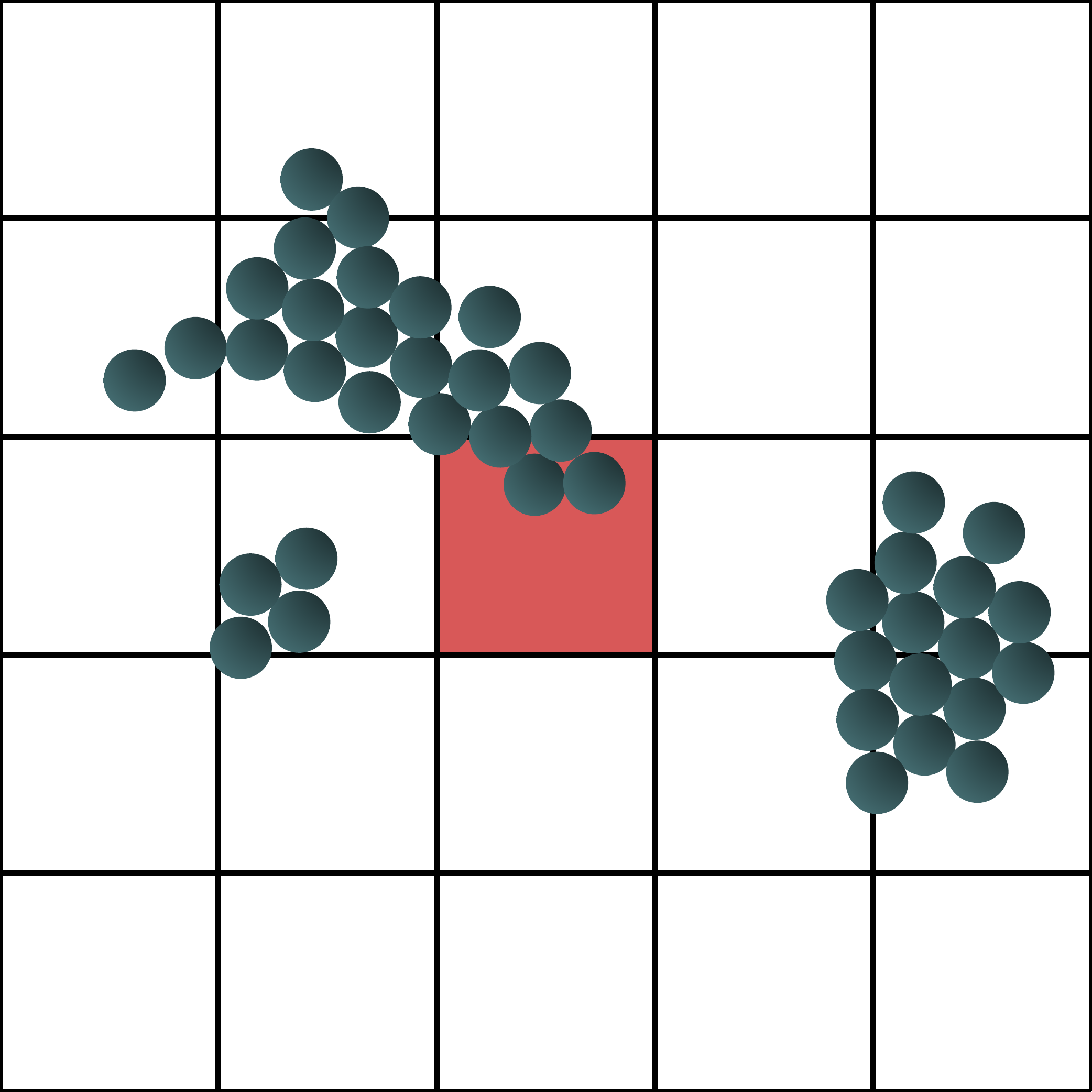}
  \includegraphics[width=0.2\linewidth]{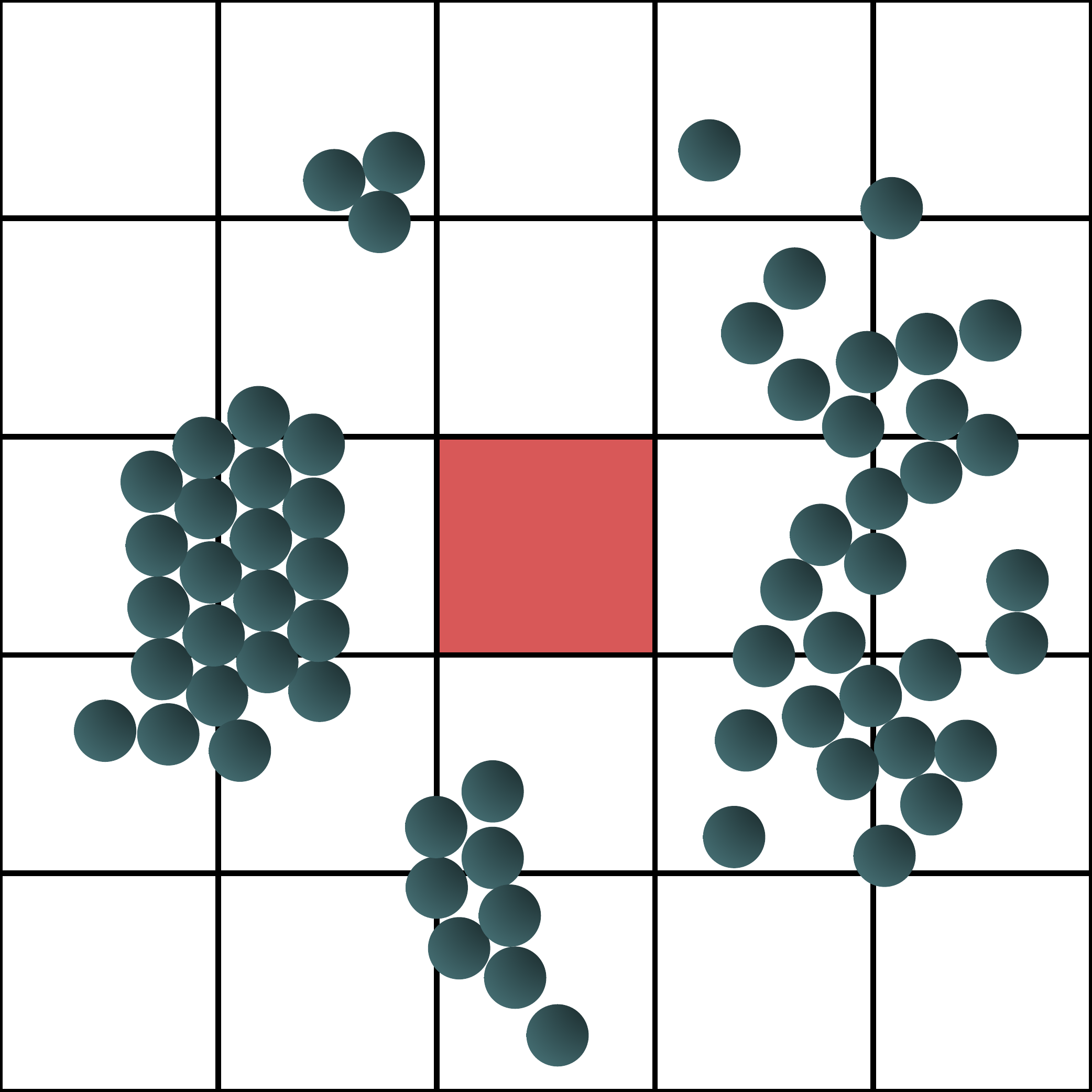}
  \includegraphics[width=0.2\linewidth]{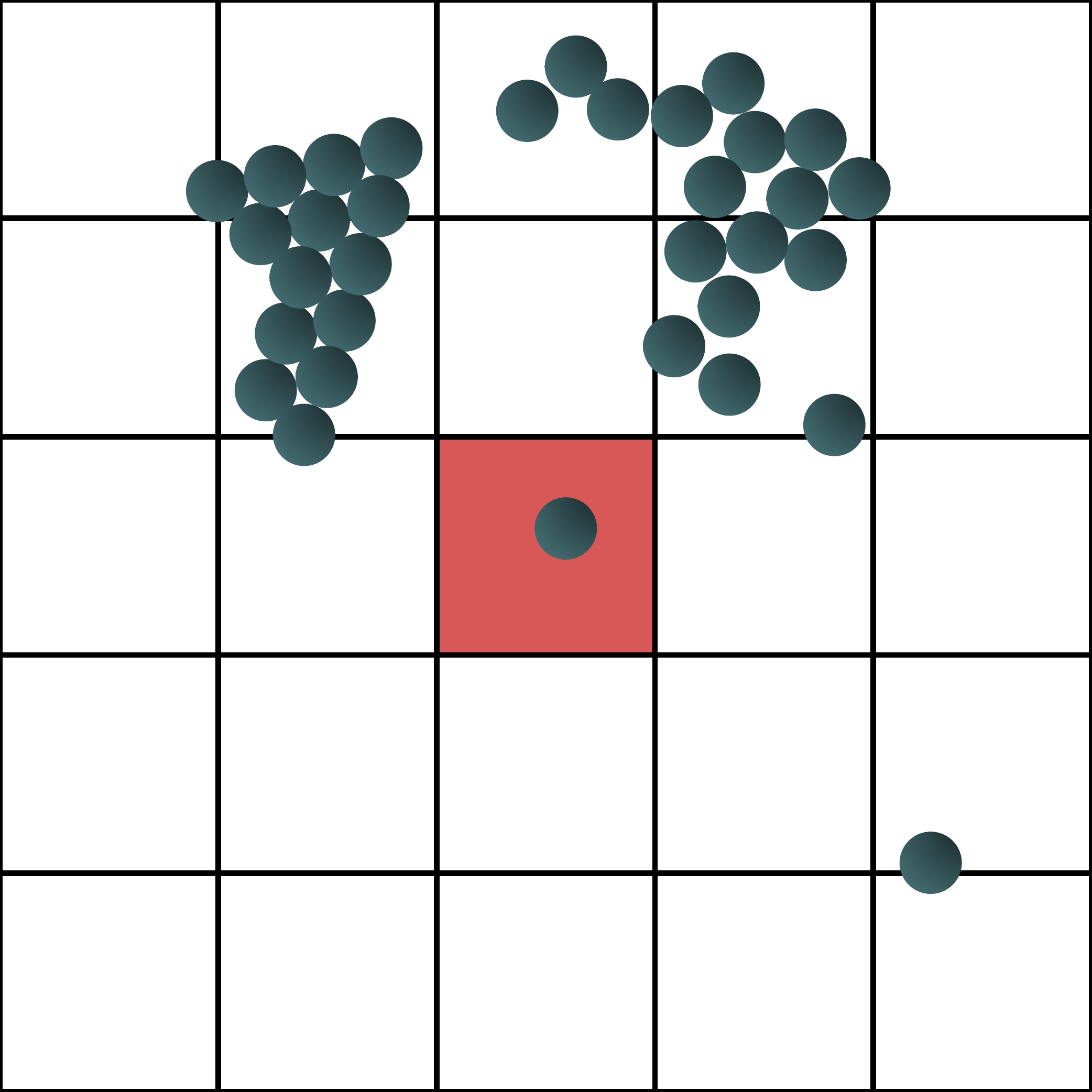}
  \includegraphics[width=0.2\linewidth]{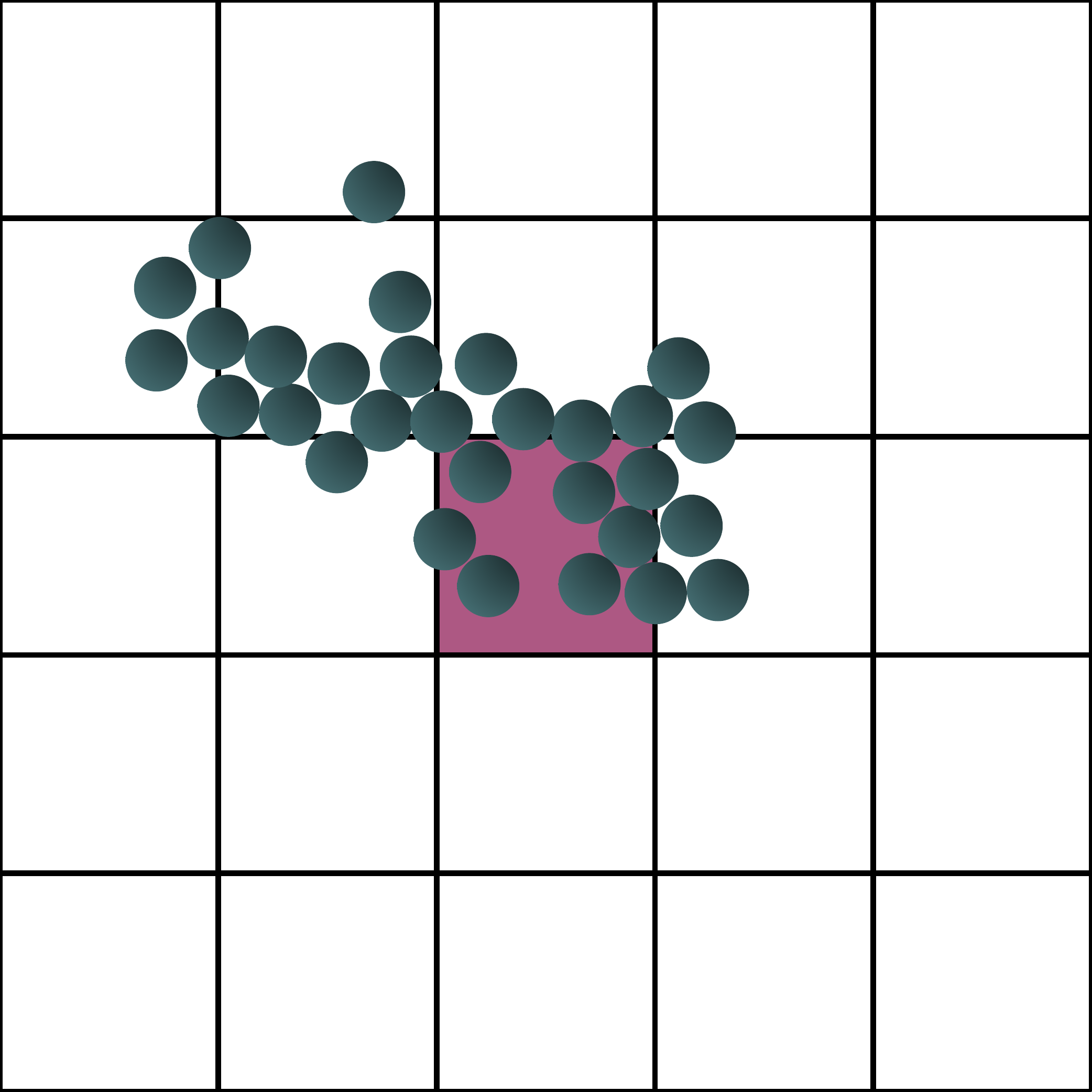}
  \includegraphics[height=0.2\linewidth]{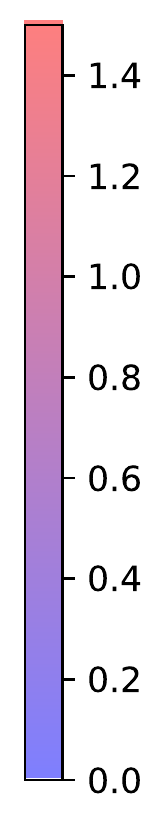}\\
  \includegraphics[width=0.2\linewidth]{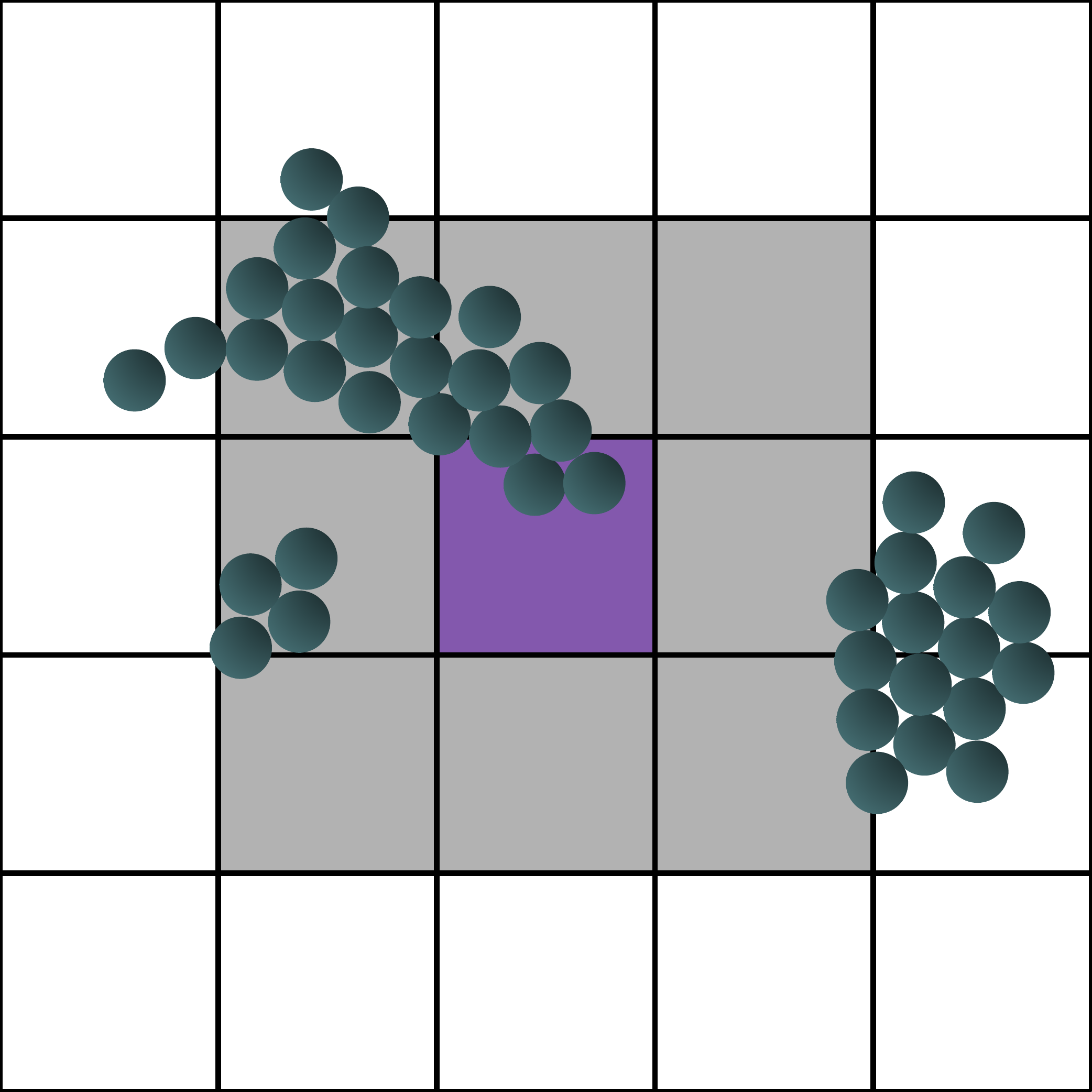}
  \includegraphics[width=0.2\linewidth]{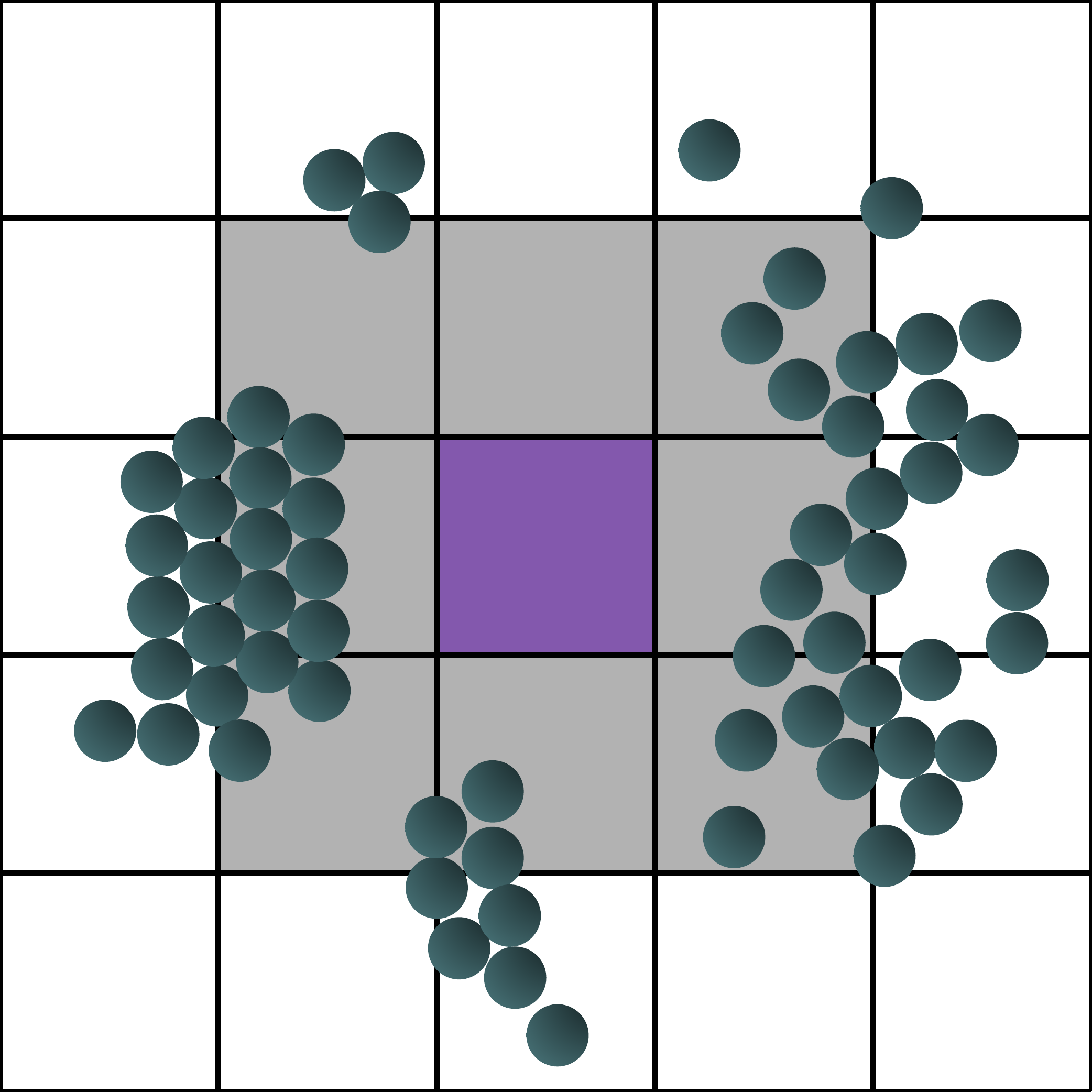}
  \includegraphics[width=0.2\linewidth]{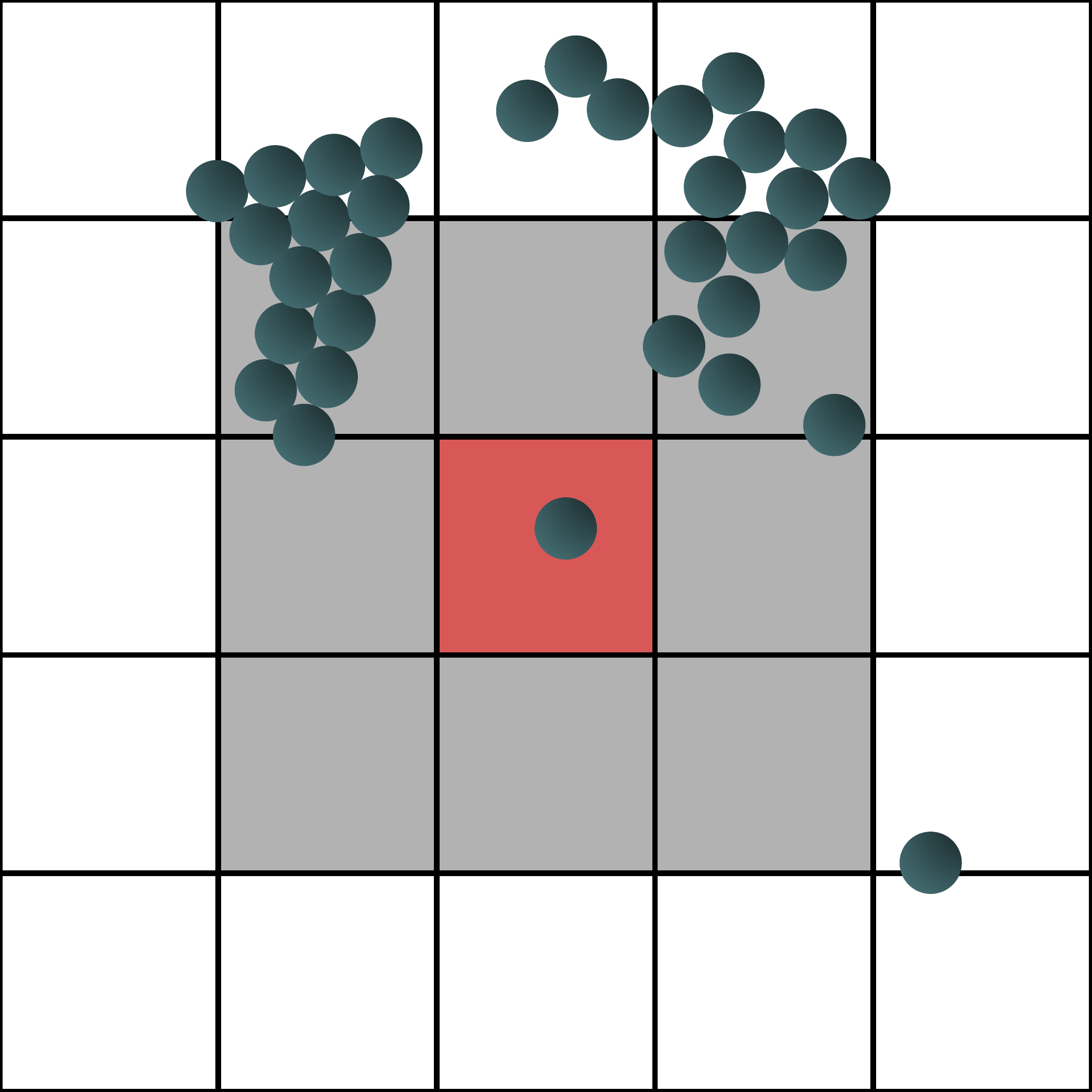}
  \includegraphics[width=0.2\linewidth]{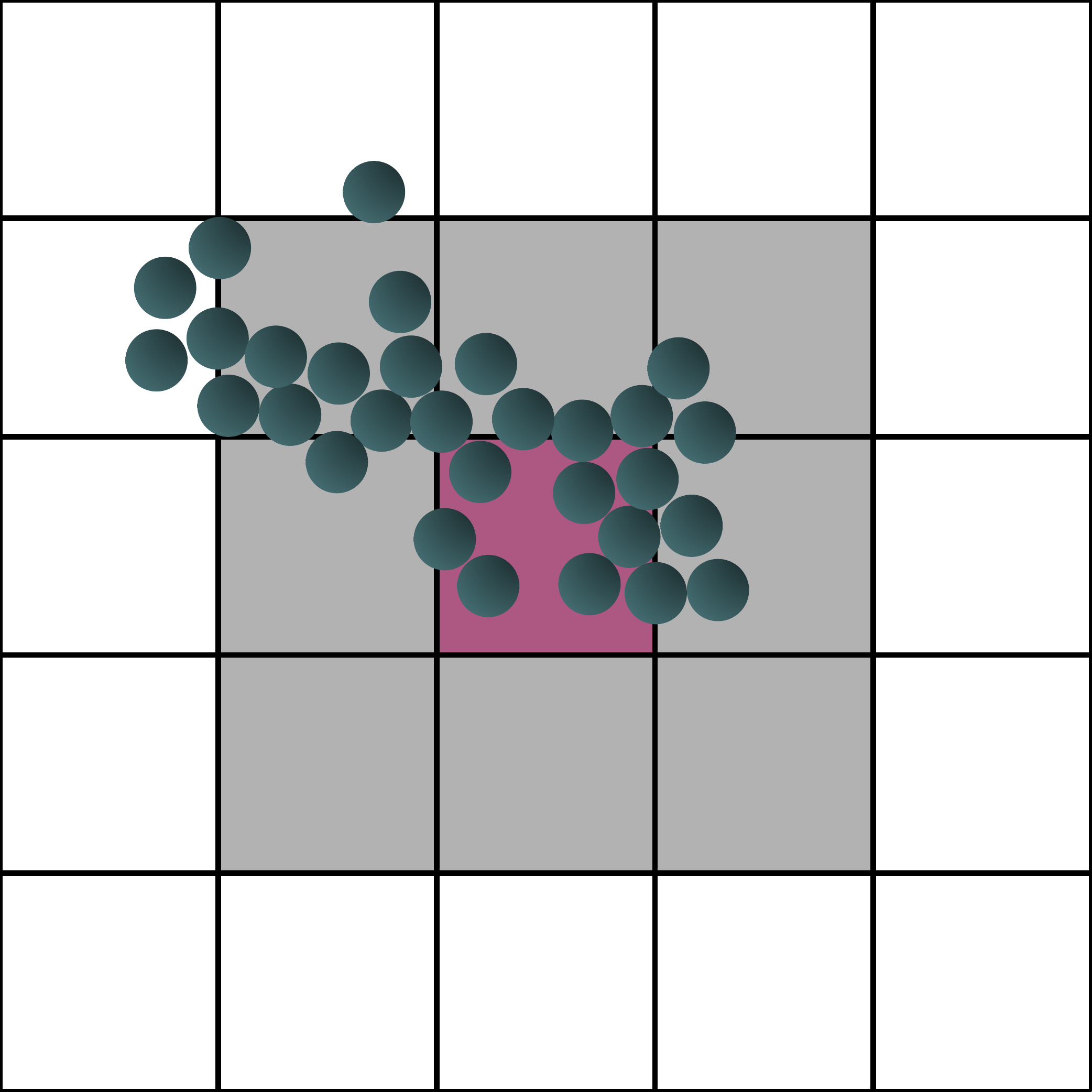}
  \includegraphics[height=0.2\linewidth]{active_colorbar.pdf}
  \end{center}
  \caption{Active Colloids. In the plots above, both the KL divergence from the instantaneous cluster size distribution to the target distribution and the absolute deviation of the mean instantaneous cluster size distribution $\mu_{\hat{\rho}(\tau)}$ from the mean target size $\mu_{\rho_{\star}}$ are shown as a function of time for the \emph{optimized} protocols of various partially decentralized approaches and a fully decentralized approach (the protocols are fully trained in these curves). In the illustrations below, we compare the predicted action in the fully decentralized case (no gray shading) and the best performing partially decentralized case, $3 \times 3_{RS}$ (with gray shading). Higher light-induced activities (red) promote cluster formation.}
  \label{fig:active}
\end{figure}

Fig.~\ref{fig:active} summarizes numerical results for a 2D colloidal particle system driven by externally controlled light~\cite{palacci_living_2013}.
In this system, when the P\'{e}clet number is sufficiently large (i.e. when the light-induced activity is high relative to the temperature), the system exhibits a nonequilibrium phase separation, called motility induced phase separation (MIPS)~\cite{cates_motility-induced_2015}.

To explore the limits of an activity-inducing external control, we sought to maintain a steady state distribution consisting of clusters of particles much smaller than the macroscopic aggregate that forms when there is constant activity. We specified a target distribution $\rho_*$ of cluster sizes using a Gaussian target distribution with $\mu=20, \sigma^2=9$ and considered a $48\times48$ grid of control.
Consistent with Prop.~\ref{prop:belief} including information of surrounding grids improves the performance of the optimal protocol. 
The best performing protocol, $3 \times 3_{CS}$, includes cluster size information and cost information in a local $3 \times 3$ region. 
Surprisingly, including just the cost information of this local region, $3 \times 3_{C}$, actually hindered the performance compared to the fully decentralized approach. 
On the other hand, for the $\textrm{Global}_C$ case, when we included information about only the cost of the entire global system, we see that the performance does remarkably well. This underscores the importance of carefully selecting relevant centralized or semi-centralized information to improve performance. 
Acting in a partially decentralized manner can not only incur an additional computation cost but can also result in a decrease in performance. 

Including surrounding information in the best performing approach was important in two cases: when the current region was empty or when it contained a cluster that was at or near the mean of the target distribution. 
In both these cases, in a fully decentralized setting, the trained agent always selects an action with the highest activity. 
However, in the partially decentralized setting, the agent enforces a lower activity when the surrounding region contains clusters that are above the mean of the target distribution, preventing the formation of a cluster far larger than the mean target size.

\section{Thermal Annealing}
\begin{figure}
  \hfill\includegraphics[width=0.9\linewidth]{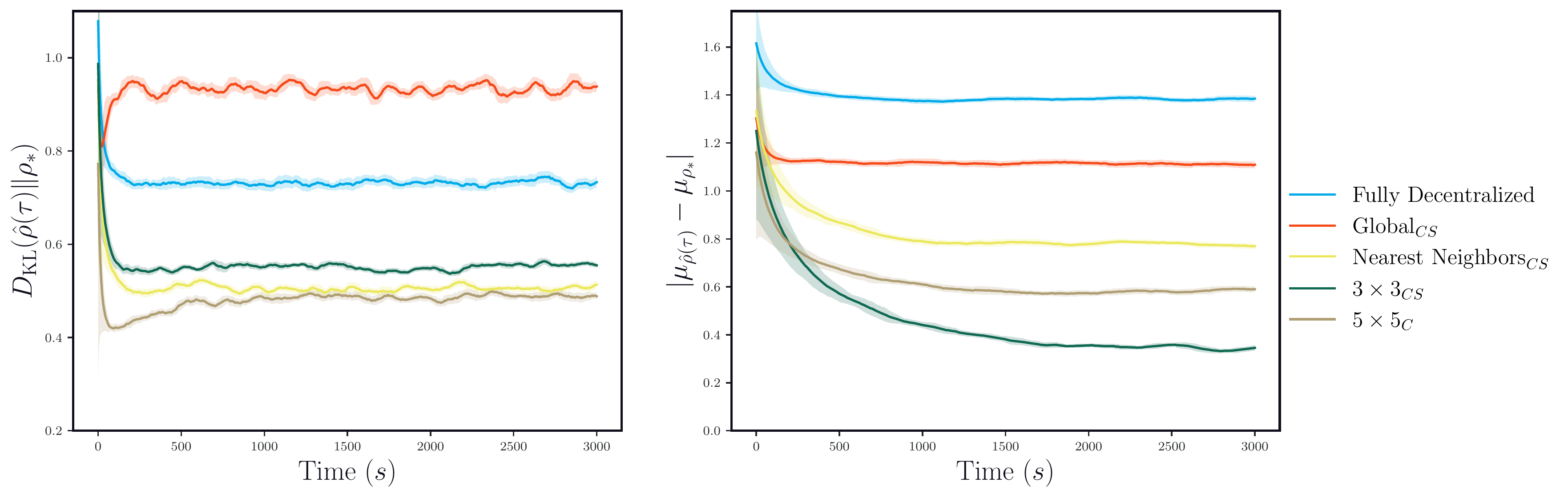}
  \begin{center}
 \includegraphics[width=0.2\linewidth]{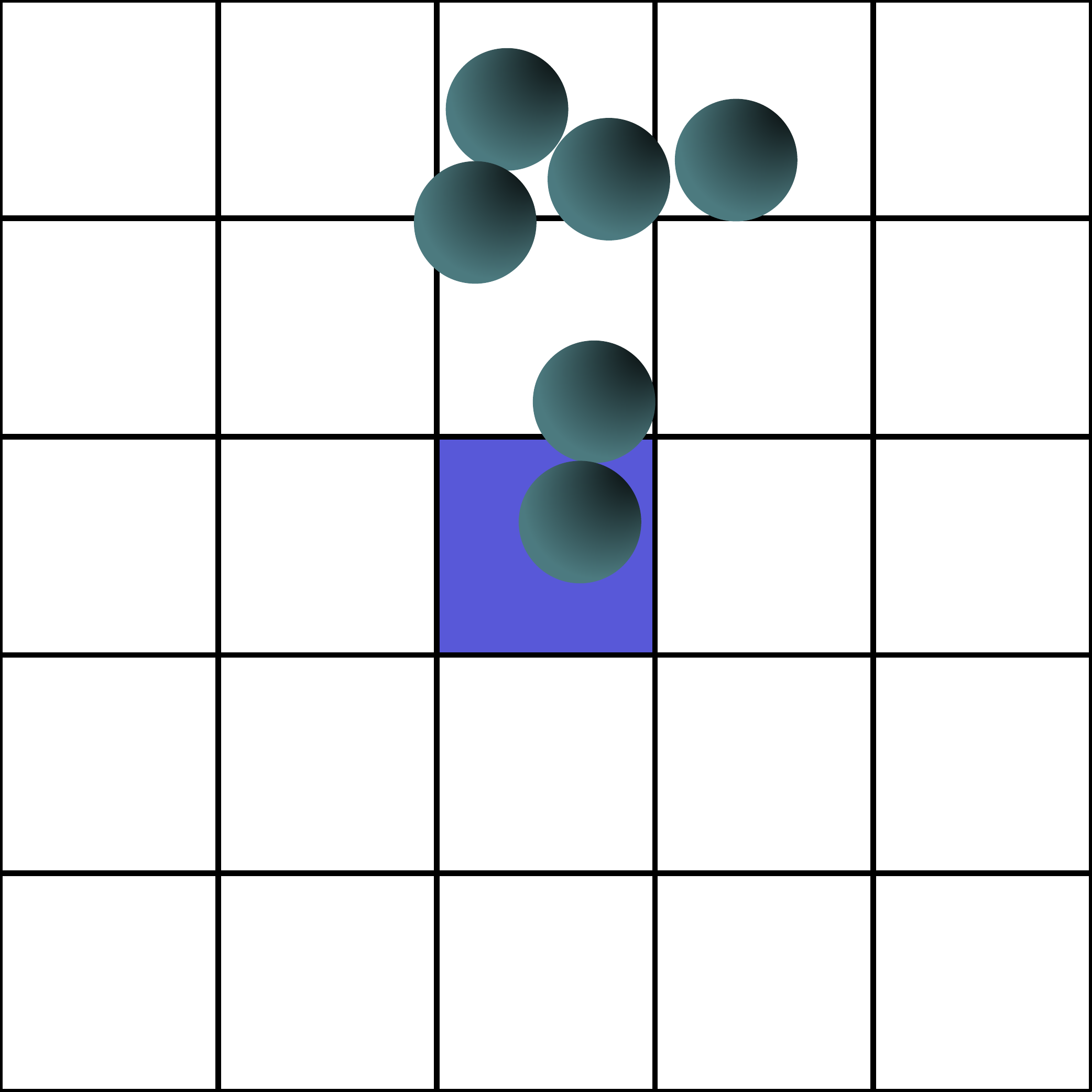}
  \includegraphics[width=0.2\linewidth]{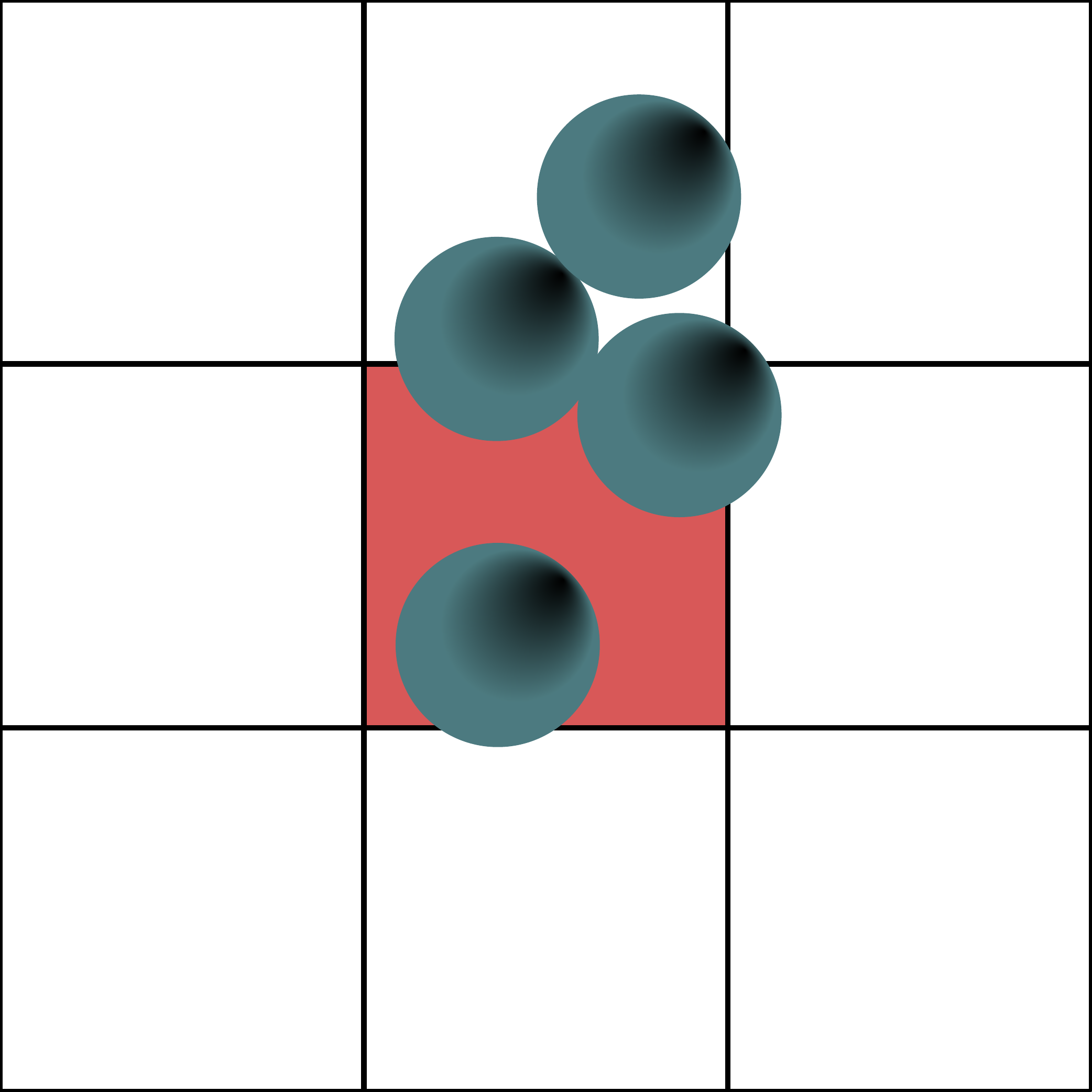}
  \includegraphics[width=0.2\linewidth]{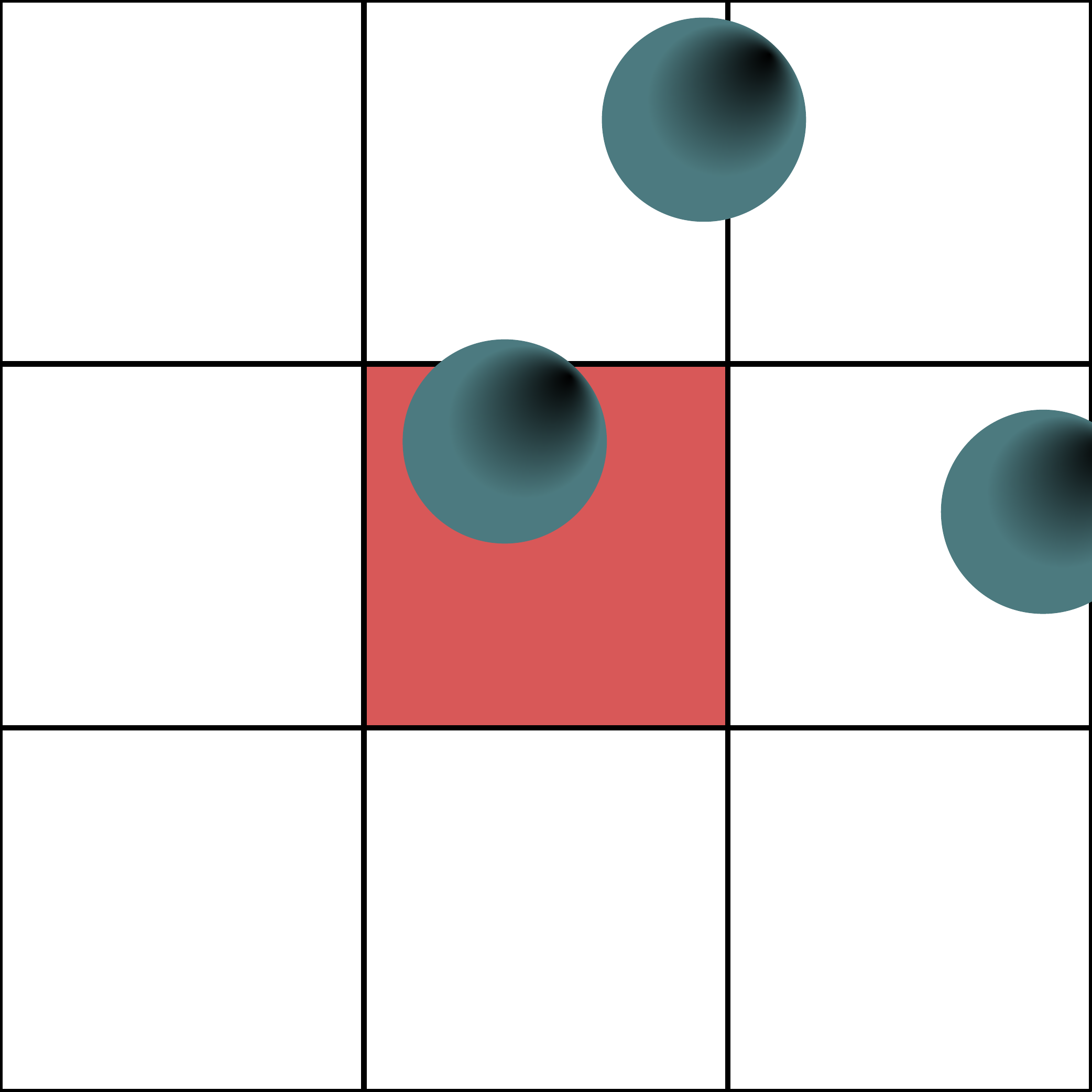}
  \includegraphics[width=0.2\linewidth]{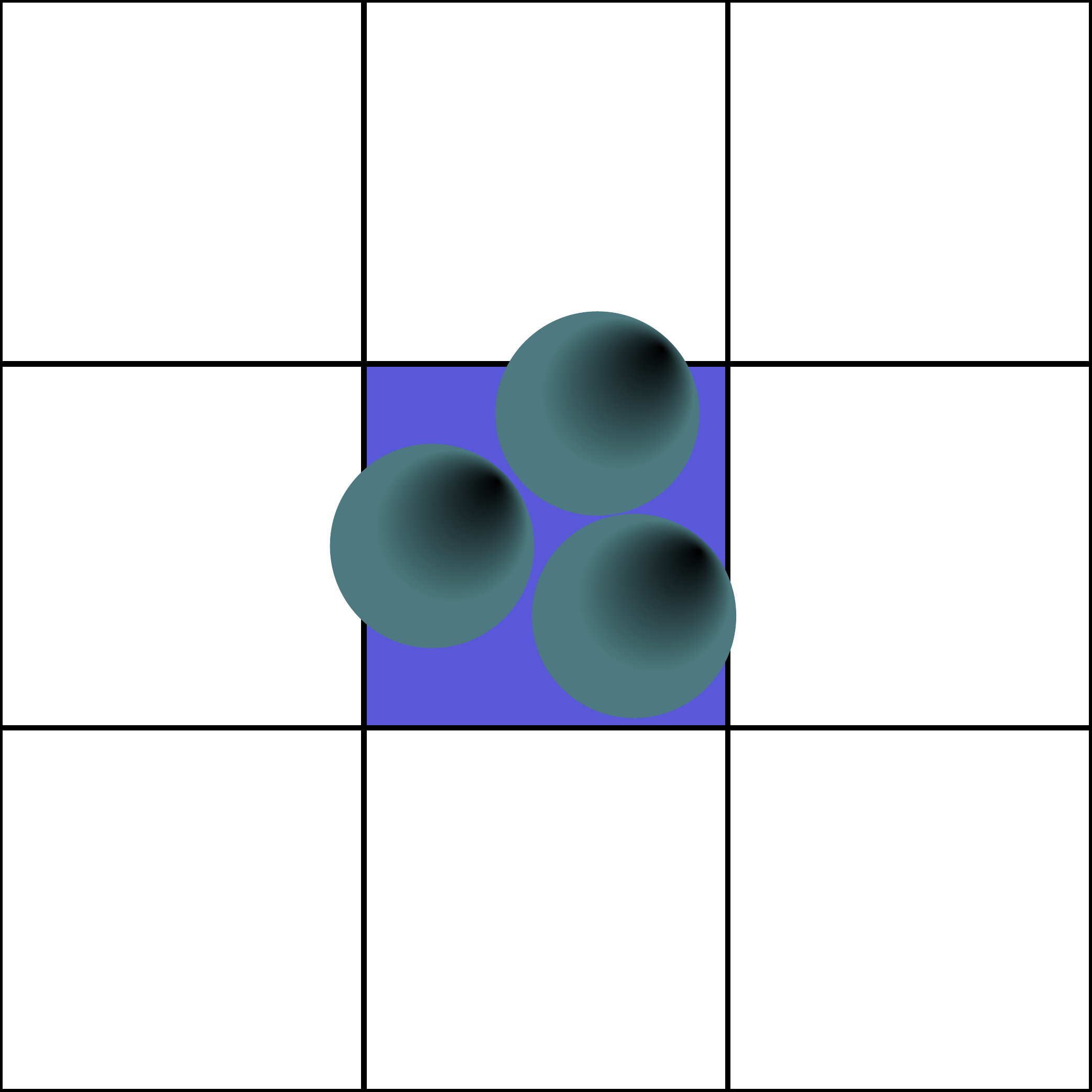}
  \includegraphics[height=0.2\linewidth]{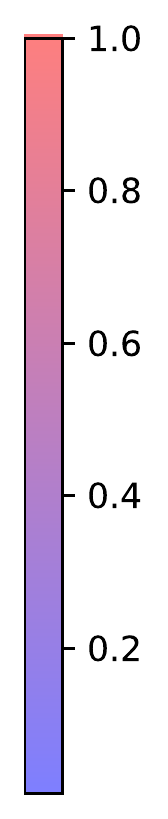}\\
  \includegraphics[width=0.2\linewidth]{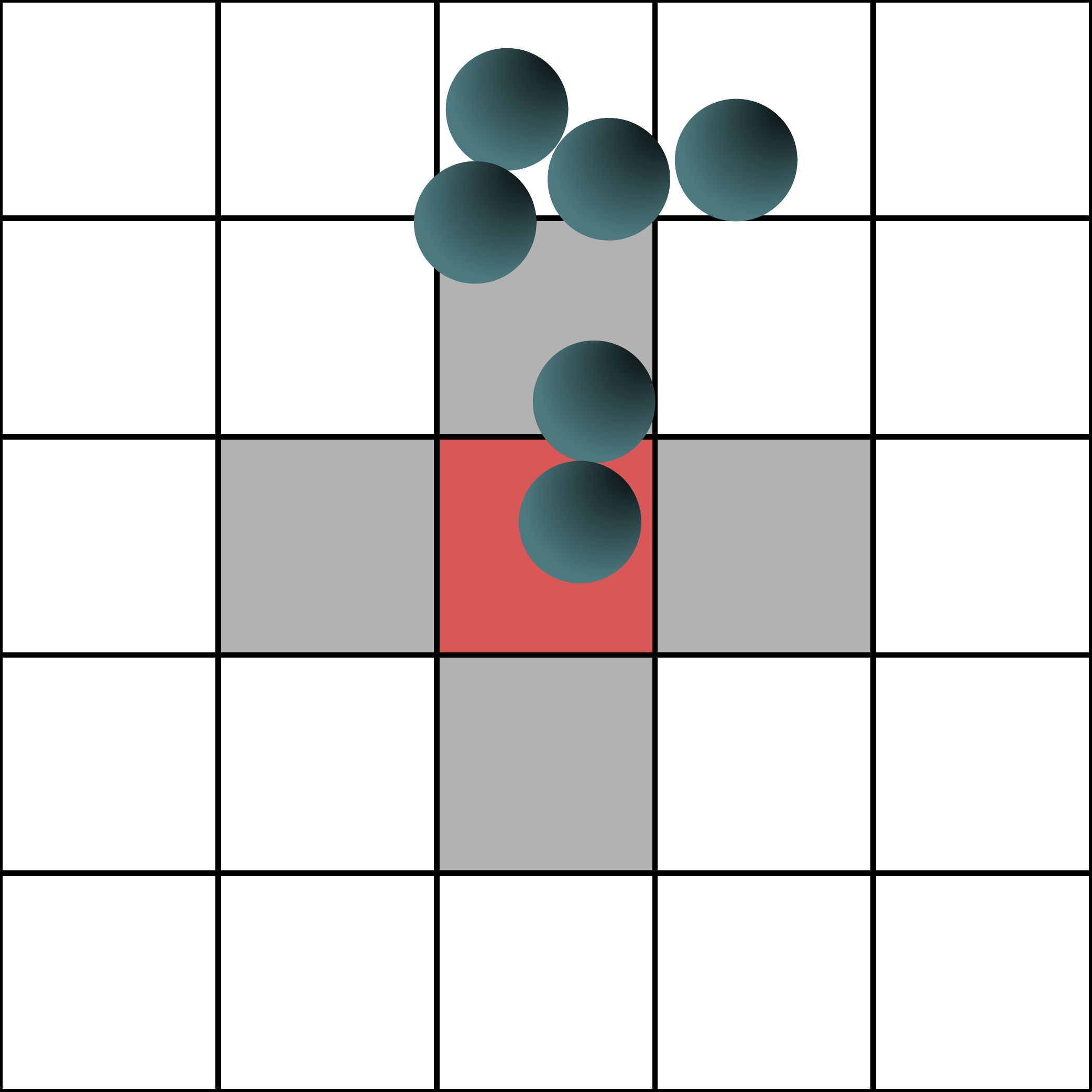}
  \includegraphics[width=0.2\linewidth]{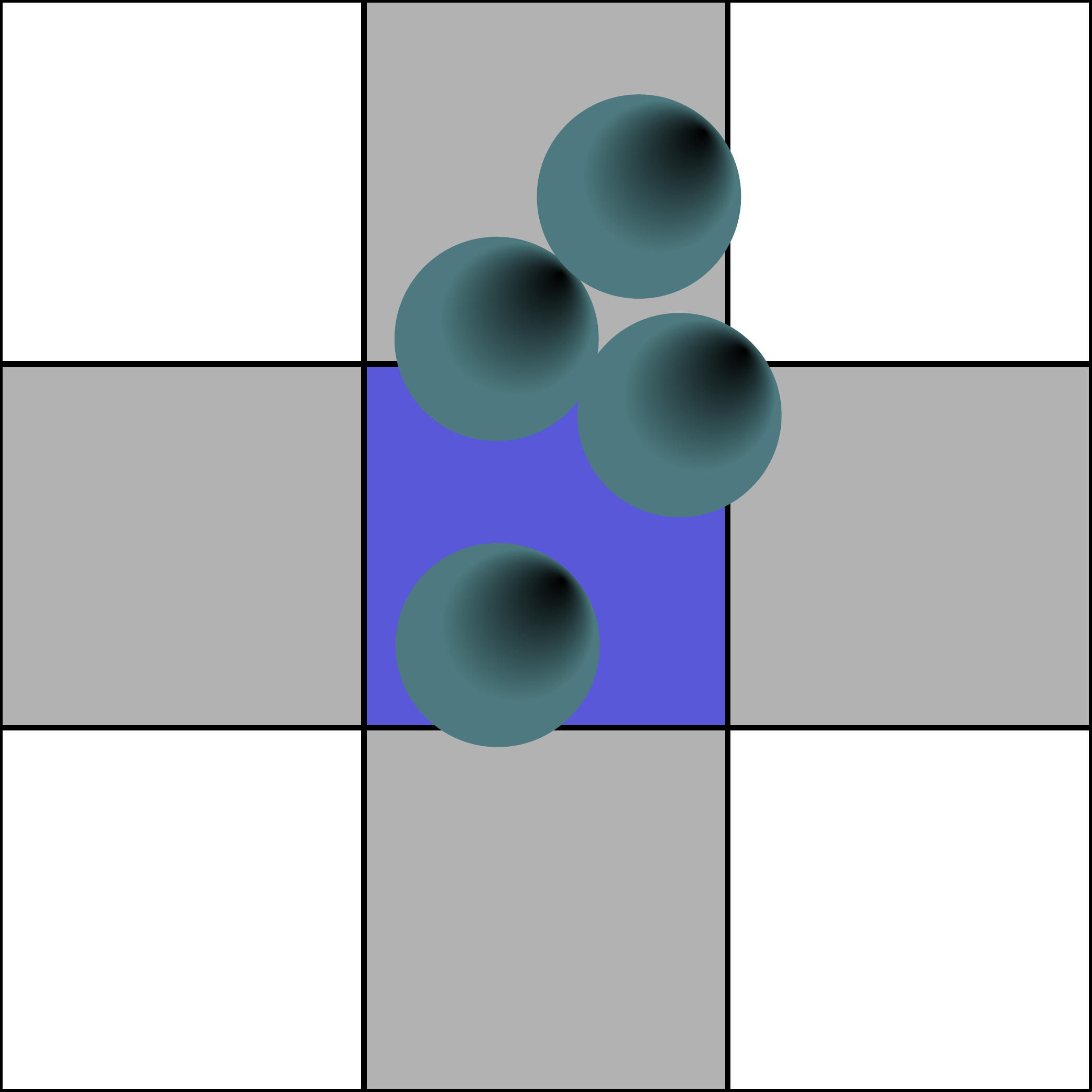}
  \includegraphics[width=0.2\linewidth]{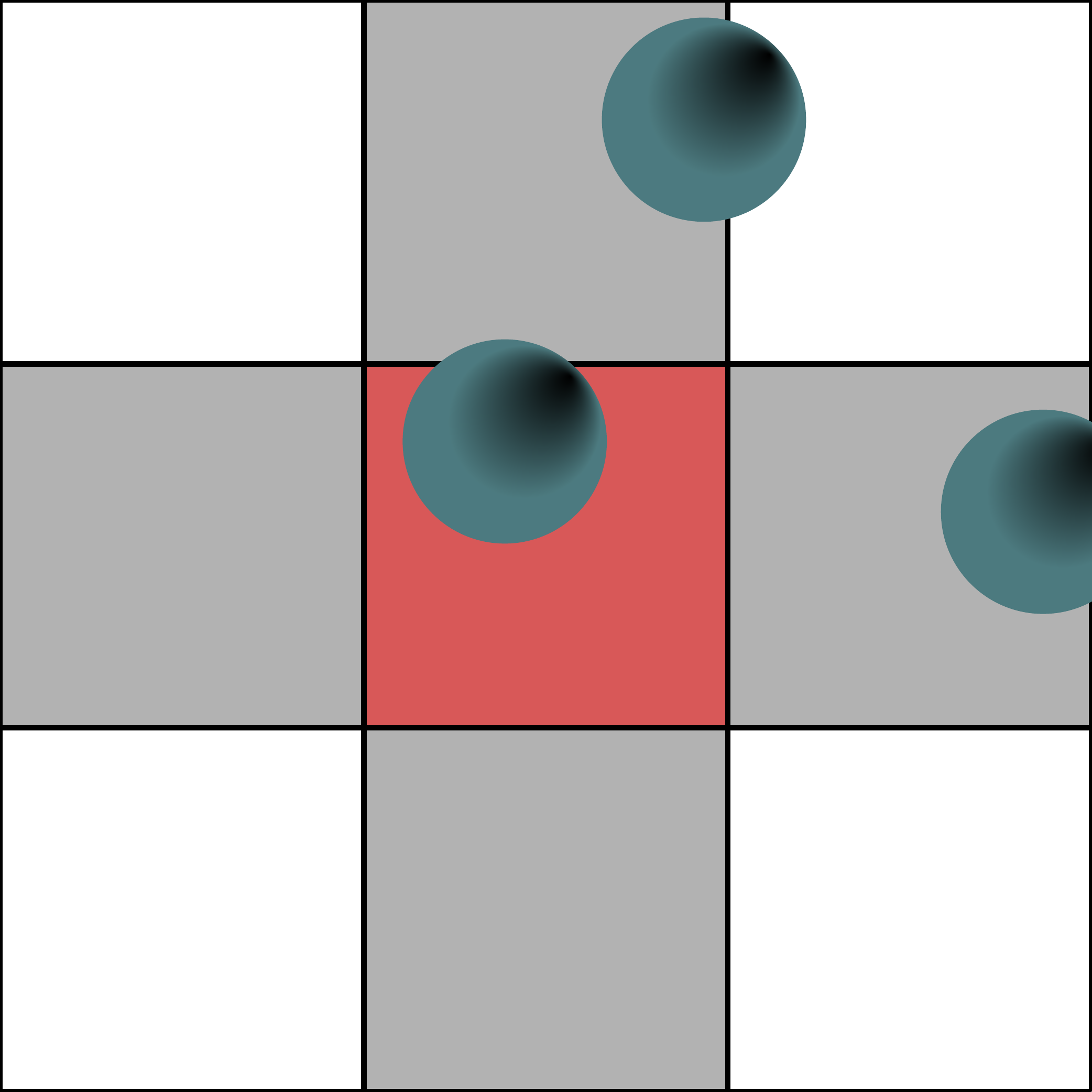}
  \includegraphics[width=0.2\linewidth]{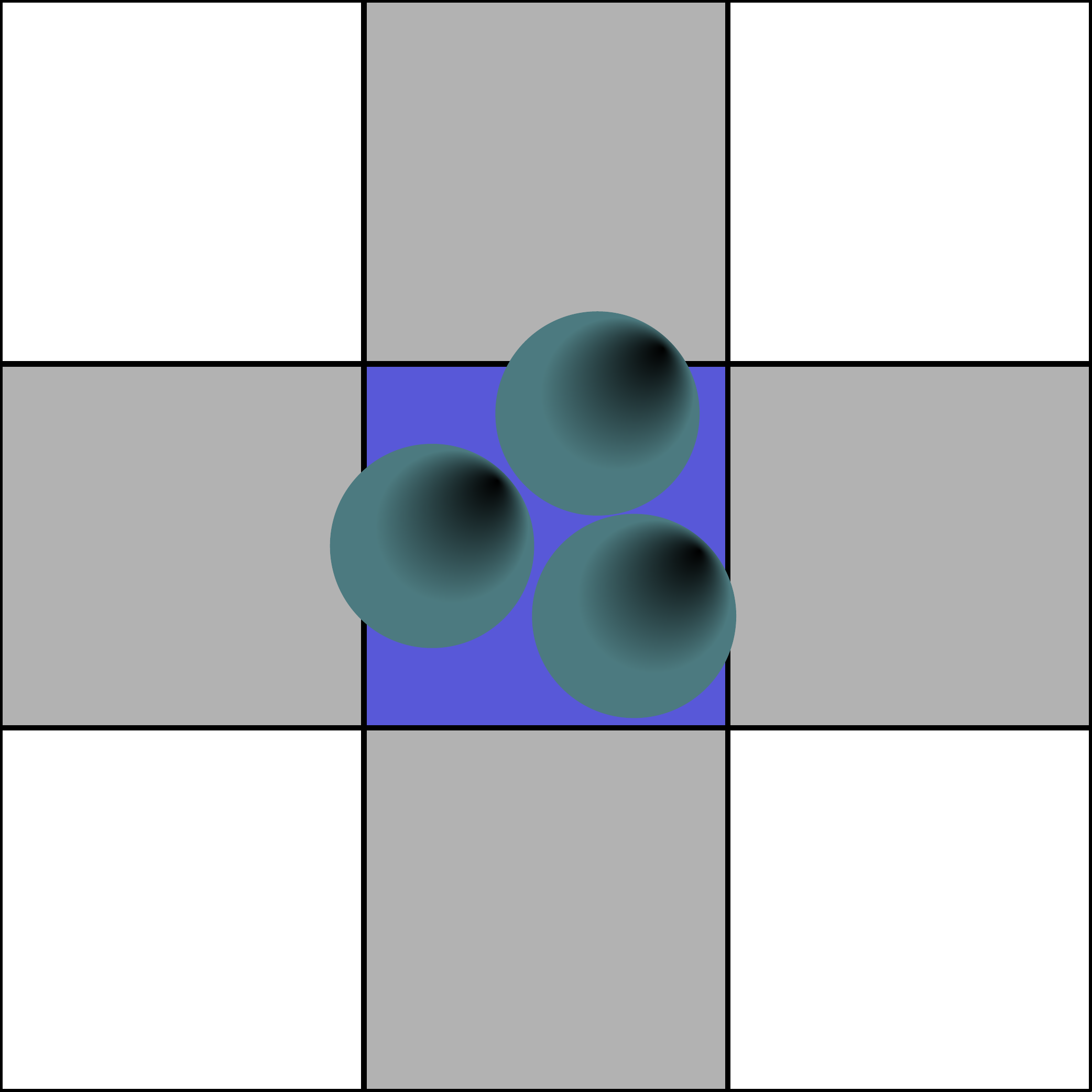}
  \includegraphics[height=0.2\linewidth]{lj_colorbar.pdf}
  \end{center}
  \caption{Thermal Annealing.
   See Fig.~\ref{fig:active} caption for descriptions of plots. In the cartoons below, we compare the predicted action in the fully decentralized case (no gray shading) and $\textrm{Nearest Neighbors}_{CS}$ (with gray shading). Lower temperatures (more blue) promote cluster formation.}
   \label{fig:anneal}
 \end{figure}

Thermal annealing is widely used to improve the yield of nanoscopic materials~\cite{makrides_temperature_2012, dey_dna_2021}. 
Annealing is limited as a mechanism for control because there is essentially only one parameter that can be tuned, the rate at which the temperature is decreased. We examine an alternative paradigm that exercises more localized control with measurement-guided feedback to design an annealing schedule.
Rather than globally tuning a temperature, we locally update the temperature on a grid, see Fig.~\ref{fig:anneal}.
As above, we fixed a target cluster size distribution, chosen to be a Gamma distribution with variance $\sigma^2=1$ and a mean $\mu=4$ and considered a 15x15 grid of control. Here, the size and density of the system is far lower and we sought to form clusters of a much smaller size compared to the active system. This allowed us to define individual grids that were only slightly larger than an individual particle, providing much finer levels of control.
Note that the interaction potential for these particles is distinct from the one that we use to model the active system and there is no nonequilibrium dynamics.
The fully decentralized and top performing partially decentralized approaches differed in their behavior when the cluster size was around the mean. In the fully decentralized case, the agent was far more conservative when a cluster size was of size 4 or greater, and would impose a high temperature breaking apart the cluster. On the other hand, in one of the top performing partially decentralized approaches, $\textrm{Nearest Neighbors}_{CS}$, the agent allowed cluster sizes between 3 and 5 to persist by ensuring a lower temperature, given that there were no clusters in the vicinity.

\section*{Broader Impact}
In this work, we investigated how a \emph{partially} decentralized reinforcement learning could be used to learn protocols that externally drive a system towards some non-equilibrium steady-state. This approach can be used as a framework for learning protocols for self-assembly in more complex systems. For example, in the design of suprastructures, each agent---or a local grouping of agents---can be used to design individual substructures, which then self-assemble to form larger structures.
At present, the key limitation is that the paradigm is most naturally suited to control with spatial resolution.
Ultimately, the ability to design protocols for self-assembly of nanoscopic structures, can have wide-ranging impacts across a number of disciplines ranging from medicine to materials science.
Potential negative impacts could, in principle, arise from the design of harmful or toxic materials.

\paragraph*{Data and Code Availability:}
The data that support the findings of this study are available from the corresponding author upon reasonable request.
Our code is available at \url{https://github.com/rotskoff-group/marl-design}.

\printbibliography 

\end{document}